\documentclass[10pt,a4paper]{article}
\usepackage[english]{babel}
\usepackage[T1]{fontenc}
\usepackage[utf8]{inputenc}
\usepackage[
	inner=2cm, 
	outer=2cm, 
	top=2.5cm, 
	bottom=2.5cm, 
]{geometry}

\usepackage{lipsum}
\usepackage{amsmath}
\usepackage{amsfonts}
\usepackage{amssymb}
\usepackage{graphicx}
\usepackage[format = plain, font = small, tablename = TABLE, figurename = FIG.]{caption}
\usepackage{subcaption}

\makeatletter
\newlength{\captionindent}
\long\def\@makecaption#1#2{%
  \vskip\abovecaptionskip
  \sbox\@tempboxa{#1: #2}%
  \ifdim \wd\@tempboxa >\hsize
    \hspace*{\captionindent}#1: #2\par
  \else
    \global \@minipagefalse
    \hb@xt@\hsize{\hfil\box\@tempboxa\hfil}%
  \fi
  \vskip\belowcaptionskip}
\makeatother

\newcommand{\pan}[1]{\left\langle#1\right\rangle}

\newcommand{\de}[1]{\!\operatorname{d}\!{#1}}
\usepackage{color}

\usepackage{float}

\usepackage[titletoc, toc, title]{appendix}

\usepackage[backend=bibtex,style=ieee,citestyle=numeric-comp,url=false,doi=false,isbn=false,date=year]{biblatex}
\title{\bfseries 
Limited data on infectious disease distribution exposes ambiguity in epidemic modeling choices
}
\author{Laura Di Domenico,$^{1,*}$ Eugenio Valdano,$^2$ and Vittoria Colizza$^{2,3}$}
\date{
\footnotesize \itshape
$^1$Institute of Social and Preventive Medicine, University of Bern, Bern, Switzerland\\
$^2$Sorbonne Université, INSERM, Institut Pierre Louis d’Epidemiologie et de Santé Publique (IPLESP), Paris, France\\ 
$^3$Department of Biology, Georgetown University, Washington, District of Columbia, USA\\
}

\usepackage{multicol}
\usepackage[autostyle=true]{csquotes}
\AtEveryBibitem{\clearlist{language}}
\AtEveryBibitem{\clearfield{pages}}
\AtEveryBibitem{\clearfield{volume}}  
\AtEveryBibitem{\clearfield{number}}
\AtEveryBibitem{\clearfield{note}}
\AtEveryBibitem{\clearfield{series}}

\renewcommand{\thesection}{\Roman{section}.}

\usepackage{titlesec}
\titleformat{\section}{\normalfont\fontsize{10}{15}\bfseries\filcenter}{\thesection}{2.0mm}{}
  
\newcommand\blfootnote[1]{%
  \begingroup
  \renewcommand\thefootnote{}\footnote{#1}%
  \addtocounter{footnote}{-1}%
  \endgroup
}
 
\begin{document}

\maketitle
\renewcommand{\abstractname}{\vspace{-\baselineskip}}
\begin{abstract}
Traditional disease transmission models assume that the infectious period is exponentially distributed with a recovery rate fixed in time and across individuals. This assumption provides analytical and computational advantages, however it is often unrealistic when compared to empirical data. Current efforts in modeling non-exponentially distributed infectious periods are either limited to special cases or lead to unsolvable models. Also, the link between empirical data (the infectious period distribution) and the modeling needs (the definition of the corresponding recovery rates) lacks a clear understanding. Here we introduce a mapping of an arbitrary distribution of infectious periods into a distribution of recovery rates. Under the markovian assumption to ensure analytical tractability, we show that the same infectious period distribution at the population level can be reproduced by two modeling schemes --~that we call \textit{host-based} and \textit{population-based}~-- depending on the individual response to the infection, and aggregated empirical data cannot easily discriminate the correct scheme. Besides being conceptually different, the two schemes also lead to different epidemic trajectories. Although sharing the same behavior close to the disease-free equilibrium, the \textit{host-based} scheme deviates from the expected epidemic when reaching the endemic equilibrium of an SIS (susceptible-infectious-susceptible) transmission model, while the \textit{population-based} scheme turns out to be equivalent to assuming a homogeneous recovery rate.  We show this through analytical computations and stochastic epidemic simulations on a contact network, using both generative network models and empirical contact data. It is therefore possible to reproduce heterogeneous infectious periods in network-based transmission models, however the resulting prevalence is sensitive to the modeling choice for the interpretation of the empirically collected data on the length of the infectious period. In absence of higher resolution data, studies should acknowledge such deviations in the epidemic predictions. 

\end{abstract}
\vspace{0.8cm}
\begin{multicols}{2}

\section{INTRODUCTION}

\blfootnote{*laura.didomenico@unibe.ch}

The infectious period has a key role in the progression of an infectious disease. It is the time interval during which an infected host can transmit the pathogen to other susceptible individuals, therefore it is closely linked to the ecological persistence of the disease and the challenges of its eradication. The infectious period depends both on disease natural history, and on the interventions possibly put in place: treatment, for instance, can be effective in reducing it. Collecting and characterizing this type of data is quite challenging, as demonstrated by statistical studies on measles \cite{bailey_statistical_1954} and long-lived infections such as HIV \cite{sabin_natural_2013}. Recent examples about COVID-19 \cite{ganyani_estimating_2020} and monkeypox \cite{miura_estimated_2022} show that a common feature of empirical infectious period distributions is their overdispersion or underdispersion, deviating from an exponential distribution. This is in contrast with the traditional analytical approach adopted in the physics community, based on modeling the infectious period with a fixed recovery rate. By this approach, each infectious host recovers at a rate $\mu$. This paradigm has a twofold advantage. First, it is easy to handle both analytically and numerically; secondly, it maps the process into a spontaneous 1-body reaction, allowing to borrow solutions from other fields of physics, like reaction-diffusion processes, and decays. However, this approach uniquely constrains the distribution of the infectious period $\tau$ to an exponential distribution having expected value $\pan{\tau}=\mu^{-1}$. Therefore, fitting $\pan{\tau}$ from real data leaves no extra degrees of freedom to model the dispersion of the data.

Efforts to overcome this problem already exist, e.g. through additional infectious compartments~\cite{anderson_spread_1980,lloyd_realistic_2001, krylova_effects_2013}, or by splitting hosts into epidemiologically relevant groups~\cite{gou_how_2017,darbon_disease_2019}. An alternative approach is to directly plug heterogeneous recovery rates into the model~\cite{de_arruda_impact_2020,bonaccorsi_epidemics_2016}. All three approaches have limitations: adding compartments limits the choice of possible distributions, partitioning individuals is limited to specific epidemiological contexts, and using distributed recovery rates is not justified by a clear link with a corresponding distribution of the infectious periods. On the other hand, parameterizing models with an explicit distribution of the infectious period makes them harder to solve. 

In this study, we develop a theory to include arbitrary distributions of infectious periods, so that they can be informed by real data. We treat both the infectious period $\tau$ and the recovery rate $\mu$ as stochastic variables, and determine the mapping between their probability distributions. We treat individual recovery as a spontaneous process occurring at a given rate $\mu$, but we sample that rate from a distribution $f$ appropriately chosen so that the resulting distribution of infectious periods $\tau$ at the population level follows a desired distribution $g$, possibly fitted on data. Our mapping allows to analytically derive $f(\mu)$ from $g(\tau)$.

This mapping also provides a clear understanding of the link between empirical data (the infectious disease distribution) and the modeling needs (the definition of the corresponding recovery rate). The infectious period distribution $g(\tau)$ is usually reconstructed from population data collected through surveillance and observational studies. The recovery rate $\mu$, instead, is a variable defined by the modeling scheme at the individual level. This implies a degeneracy of models that may assign individual rates differently, but produce the same $g(\tau)$. We study this by defining the {\itshape host-based} scheme, by which each host is given a recovery rate $\mu$ that is fixed in time and sampled from $f$, and the {\itshape population-based} scheme, by which every time a host recovers it re-samples its recovery rate from $f$. The latter scheme models a scenario in which the chance of recovery for a specific host changes after each re-infection, for instance, because of a difference in the immunity response. We show that both schemes recover the same distribution $g$ computed from $f$ through our mapping. However, we prove that, while the two schemes share their critical behavior (close to the epidemic threshold), they significantly differ in the endemic equilibrium. This has far-fetching implications, as the choice of the correct scheme may be nonunivocal, depending on the epidemic context, but data collection often does not allow to empirically discriminate between the {\itshape host-based} and the {\itshape population-based} schemes.

In Sec.~\ref{sec:mapping}, we build and discuss the analytical mapping from $g$ to $f$. We also define the {\itshape host-based} and {\itshape population-based} schemes. We then characterize the epidemic threshold (Sec.~\ref{sec:epithreshold}) and the endemic equilibrium (Sec.~\ref{sec:endemic}).
In Sec.~\ref{sec:casestudy}, we apply our methodology to the spread of nosocomial infections in healthcare settings using empirical contact data. In Sec.~\ref{sec:conclusion} we discuss the implications of our results for the modeling community.

\section{MAPPING INFECTIOUS PERIODS INTO RECOVERY RATES}
\label{sec:mapping}
We consider the Susceptible-Infected-Susceptible (SIS) model \cite{kermack_contribution_1927, anderson_infectious_1992}.
A susceptible individual becomes infected at rate $\lambda$ upon contact with an infected host. It then recovers spontaneously to the susceptible state at rate $\mu$. Recovery confers no immunity. The transmission rate $\lambda$ is constant, while the recovery rate $\mu$ is a stochastic variable with distribution $f$. A varying $\lambda$ with constant $\mu$ was studied in Ref.~\cite{starnini_equivalence_2017} for other purposes. Let $\tau$ be the stochastic variable representing the infectious period, with distribution $g$. In the case of a fixed $\mu$, the distribution $g$ would be the exponential distribution.

Just as in a standard reaction-diffusion process, recovery is Markovian once the recovery rate is fixed, i.e. $\tau|\mu \sim Exp(\mu)$. Under this assumption, we can write 
\begin{equation}
    g(\tau)= \int_{0}^{\infty} \de{\mu} f(\mu) \mu e^{-\mu\tau} = -\frac{\de{}}{\de{\tau}}\mathcal{L}[f](\tau),
\label{eq:joint}
\end{equation}
where $\mathcal{L}$ is the Laplace transform operator. By integrating in $\tau$ and solving for $f$, we obtain   
\begin{equation}
    f(\mu) = \mathcal{L}^{-1}\left[ \hat{G} \right](\mu),
\label{eq:laplace}
\end{equation}
where $\hat{G}$ is the tail distribution function of $\tau$, i.e. $\hat{G}(\tau) = \int_{\tau}^{\infty} \de{x}\,g(x)$.
Eq.~(\ref{eq:laplace}) is solvable either by explicit computation of the inverse Laplace transform --~when possible~-- or by numerical integration~\cite{walker_laplace_2017}.
In the latter case, it is possible to determine beforehand if the solution exists by noticing that one can generate the moments of $f$ by repeatedly deriving Eq.~(\ref{eq:joint}) with respect to $\tau$, and then setting $\tau=0$:
\begin{align}
    m_0 &= 1; \\
    m_{n} &= (-1)^{n-1} \left. \frac{\de{}^{n-1}}{\de{\tau}^{n-1}}g(\tau) \right|_{\tau=0} \qquad \forall n\in \mathbb{N} \setminus \left\{0\right\},
\end{align}
where $m_n$ is the $n$-th moment of $f$, i.e. $m_n = \int_{0}^{\infty} \de{\mu} \mu^n f(\mu)$. As such, determining if the solution of Eq.~(\ref{eq:laplace}) exists maps onto a Stiltjes moment problem \cite{schmudgen_moment_2017} (see Appendix~\ref{sec:momentproblem}).

Table~\ref{table:conversion} reports the expression of $f(\mu)$ for some commonly used distributions of infectious periods: exponential, gamma-distributed, power-law.

The mapping introduced above works for both the {\itshape host-based} and {\itshape population-based} schemes. The difference relies on the fact that in the former scheme each host samples its $\mu$ from $f$ only once at the beginning, while in the latter each host re-samples its $\mu$ every time it recovers. Therefore, in terms of infectious period, in the {\itshape population-based} scheme each individual follows the same infectious period distribution $g(\tau)$, while in the {\itshape host-based} scheme each individual is characterized by a different (exponential) infectious period distribution, producing the distribution $g(\tau)$ when aggregated at the population level.

\section{EPIDEMIC THRESHOLD}
\label{sec:epithreshold}

The epidemic threshold is the critical value $\lambda_c$ of the transmission rate that discriminates between the disease-free state ($\lambda < \lambda_c$), and the endemic regime ($\lambda > \lambda_c$). The computation of the epidemic threshold provides an important public health metric to evaluate intervention policies~\cite{darbon_disease_2019, valdano_reorganization_2021}. 

The epidemic threshold depends on both disease features (transmission, recovery), and the topology of the underlying network of contacts along which the spreading occurs. We assume a network of $N$ nodes with adjacency matrix $A$. Let $x_{i}(t)$ be the probability that node $i$ is infectious at time $t$, with $i=1,\dots,N$. In the {\itshape host-based} scheme, node $i$ has a fixed recovery rate $\mu_i$. Following the microscopic Markov chain formalism~\cite{chakrabarti_epidemic_2008, castellano_thresholds_2010, gomez_discrete-time_2010}, we can write the differential equations describing the evolution of the disease as a perturbation of the disease-free state (thus neglecting $\mathcal{O}\left(x_i x_j\right)$ and higher orders):
\begin{equation}
    \frac{\de x_i(t)}{\de t} = -\mu_i x_i(t) + \lambda \sum_{j} A_{ij} x_j(t)
\label{eq:diff}
\end{equation}
In matrix form this reads $\dot{\bf{x}}=(-M +\lambda A)\textbf{x}$, where $\textbf{x}=(x_{1}(t), \dots,x_{N}(t))$ and $M=\mbox{diag}\{\mu_1,\dots,\mu_N\}$ is the diagonal matrix containing all the recovery rates, which have been sampled from $f(\mu)$ at $t=0$.
The epidemic threshold $\lambda_c$ then solves the equation
\begin{equation}
 \rho(-M + \lambda_c A) = 0
 \label{eq:sogliaeq}
\end{equation}
as proven in \cite{wang_epidemic_2003, gomez_discrete-time_2010}, where $\rho$ indicates the spectral radius, i.e. the largest eigenvalue. 

In the {\itshape population-based} scheme, we can observe that, close to the disease-free equilibrium, re-infection events are suppressed and can thus be dropped in the threshold computation as higher-order terms. This means that we can neglect the update mechanism of $\mu$, and retrieve the same epidemic threshold as in the {\itshape host-based} scheme.

In the standard case of exponentially distributed $\tau$ (i.e., constant $\mu$), Eq.~(\ref{eq:sogliaeq}) reduces to 
\begin{equation}
    \lambda_c = \frac{\mu}{\rho(A)} = \frac{1}{\pan{\tau} \rho(A)},
    \label{eq:thr_homo}
\end{equation}
given that the rate of the exponential distribution coincides with the inverse of its expected value. We now solve Eq.~(\ref{eq:sogliaeq}) in the case of a non-exponential distribution $g(\tau)$, i.e. with heterogeneous recovery rates $\mu_i$. In practical applications, the matrix $A$ often comes from a generative network model designed to reproduce key topological features of the contact structure of the population under study. Ref.~\cite{valdano_exact_2019} argues that generative network models are representable in terms of adjacency matrices whose rank equals the number of node features constrained. For instance, if one just fixes the expected degree of each node (the so-called configuration model~\cite{pastor-satorras_epidemic_2001,boguna_absence_2003,boguna_langevin_2009, pastor-satorras_epidemic_2015}), one will get the rank-$1$ adjacency matrix $A=K K^T / \left(N \pan{k}\right)$, where $K$ is the $N$-dimensional vector containing the expected degree of each node, and $\pan{k}$ is the average expected degree. For the generic rank-$r$ model one can write 
\begin{equation}
    A = V \Delta V^{T}
    \label{matrix_representation}
\end{equation}
where $V$ is an $N\times r$ matrix encoding node properties, and $\Delta$ is a $r$-dimensional bilinear form encoding the geometry of the model (see Ref.~\cite{valdano_exact_2019}). The epidemic threshold of the generic network model requires plugging Eq.~(\ref{matrix_representation}) in Eq.~(\ref{eq:sogliaeq}), and working out the calculations under the assumption that the node properties fixed by the model are uncorrelated with recovery rates (see Appendix~\ref{sec:epithresholdApp} for explicit computations). Unexpectedly, this leads to the following expression of the epidemic threshold
\begin{equation}
    \lambda_c = \frac{1}{ \pan{\tau} \rho(A)},
    \label{gen_threshold}
\end{equation}
which is the same as Eq.~(\ref{eq:thr_homo}), when expressed in terms of the average infectious period. This result shows that only the average infectious period impacts the epidemic threshold. The distribution $g(\tau)$ may be arbitrarily complex, but its first moment is enough to discriminate between disease extinction and endemicity. A model with fixed $\mu$ is therefore sufficient to study the critical behavior of disease spreading, and this is beneficial in two aspects: i) such a model is analytically and numerically the simplest possible, ii) estimating $\pan{\tau}$ from data is easier than fitting the full distribution, especially if the available sample is small. However, Eq.~(\ref{gen_threshold}) also warns us about the misuse of the recovery rate. It rigorously proves that the relevant observable is indeed the average infectious period, and not the average recovery rate. Replacing $1/\pan{\tau}$  with $\pan{\mu}$ in Eq.~(\ref{eq:thr_homo}) would lead to an overestimation of the epidemic threshold, because the identity
\begin{equation}
    \pan{\tau^j} = j! \pan{\mu^{-j}}, \; \forall j\in\mathbb{N}
    \label{eq:relazMomenti}
\end{equation}
(provable from Eq.~(\ref{eq:joint})), combined with Jensen's inequality~\cite{jensen_sur_1906}, implies that $\pan{\mu} \geq \pan{\tau}^{-1}$. Overestimating the epidemic threshold is potentially harmful, as it leads to underestimation of the risk of the disease becoming endemic.

\section{ENDEMIC PREVALENCE}
\label{sec:endemic}

Above the epidemic threshold, the SIS model converges to an endemic equilibrium characterized by a certain disease prevalence (i.e., fraction of population infected at a given time)~\cite{keeling_modeling_2007}. Computing this quantity completes the epidemiological characterization of the SIS epidemic. Quantifying the endemic prevalence, alongside the epidemic threshold, is relevant from a public health perspective, as it allows to anticipate the impact of the disease spreading in the long-term. 

The endemic equilibrium is typically harder to derive analytically concerning the epidemic threshold: no closed-form solution exists beyond homogeneous mixing even in the case of exponentially distributed $\tau$. Here we focus on homogeneous mixing, i.e. a sequence of Erdős–Rényi networks~\cite{erdos_random_1959}, and compute the corresponding endemic prevalence in the {\itshape population-based} and {\itshape host-based} scheme. Appendix~\ref{sec:endeq} contains a generalization to the configuration model for the {\itshape population-based} scheme.

To proceed, we divide compartments $I$ and $S$ into sub-classes according to the recovery rate. Classes $I_j(t)$ and $S_{j}(t)$ represent, respectively, the number of infected and susceptible individuals at time $t$ with recovery rate equal to $\mu_j$. Recovery rate thus gets formally discretized in an arbitrarily large number of values.
The spreading equations are
\begin{equation}
    \frac{\de{}}{\de{t}} I_j(t) = -\mu_j I_j(t) + \lambda \frac{1}{N} S_j(t) \sum_h I_h(t)
\label{eq:pop} 
\end{equation}
where the average connectivity of the homogeneous network is absorbed in the transmission rate $\lambda$. At time $t=0$, we have a fraction $f(\mu_j) \de{\mu_j}$ of the total number of individuals $N$ that have rate $\mu_j$.
In the {\itshape population-based} scheme, as time passes, large values of $\mu$ are replaced sooner, as they generate, on average, shorter infectious periods.
Likewise, smaller values of $\mu$ persist longer. This implies that the fraction of hosts with rate $\mu_j$ at a given time deviates from the initial fraction $f(\mu_j)\,\de{\mu_j}$ as time passes.
Notwithstanding, if we look exclusively at compartment $S_{j}(t)$, we can state that $S_{j}(t)=f(\mu_j) \de{\mu_j}\,S(t)$, because a new $\mu_j$ is assigned after recovery, and the inter-event time between recovery and re-infection does not depend on the recovery rate $\mu_j$ of the susceptible individual.
By inserting this in  Eq.~(\ref{eq:pop}), we can then set the {\itshape rhs} equal to zero and sum over $j$  to get rid of the discretization and obtain the endemic prevalence at equilibrium
\begin{equation}
   x_{eq} = \frac{I_{eq}}{N} = 1 - \frac{1}{\pan{\tau} \lambda}.
   \label{eq:endemicHOMO}
\end{equation}
So we find that, as for the epidemic threshold, the endemic equilibrium in the {\itshape population-based} scheme depends only on the average infectious period, regardless of its distribution, and coincides with the equilibrium obtained assuming a homogeneous recovery rate $\mu = \pan{\tau}^{-1}$. 

It is a different matter for the {\itshape host-based} scheme. 
We find a new formula to analytically derive endemic prevalence $x_{eq}$ for homogeneous mixing, as a solution of the following  equation:

\begin{equation}
    \mathcal{L}\left[g\right]\left( \lambda x_{eq} \right) = 1-x_{eq}.
    \label{eq:endemicL}
\end{equation}
Details of the derivation can be found in Appendix~\ref{sec:endEqui}.
In general Eq.~(\ref{eq:endemicL}) is solvable numerically. In some cases it leads to an analytic expression for $x_{eq}$. One of such cases is obviously when $\tau$ is exponentially distributed, giving the same result as in Eq.~(\ref{eq:endemicHOMO}). If $\tau$ is gamma-distributed (see Tab.~\ref{table:conversion}), Eq.~(\ref{eq:endemicL}) becomes relatively simple: $\left( 1+x_{eq}\lambda\pan{\tau}/\kappa \right)^{-\kappa}=1-x_{eq}$, where we used the parameterization $\pan{\tau}=\kappa\theta$. Then, further assuming $\kappa=1/2$, gives
$x_{eq} = 1 - \frac{1+\sqrt{1+8 \pan{\tau} \lambda}}{4 \pan{\tau}\lambda}$.
This example explicitly shows how different the endemic equilibrium can be from the exponentially-distributed case (Eq.~(\ref{eq:endemicHOMO})).

\begin{figure*}[htbp]   
  \centering
  \includegraphics[width=0.75\linewidth]{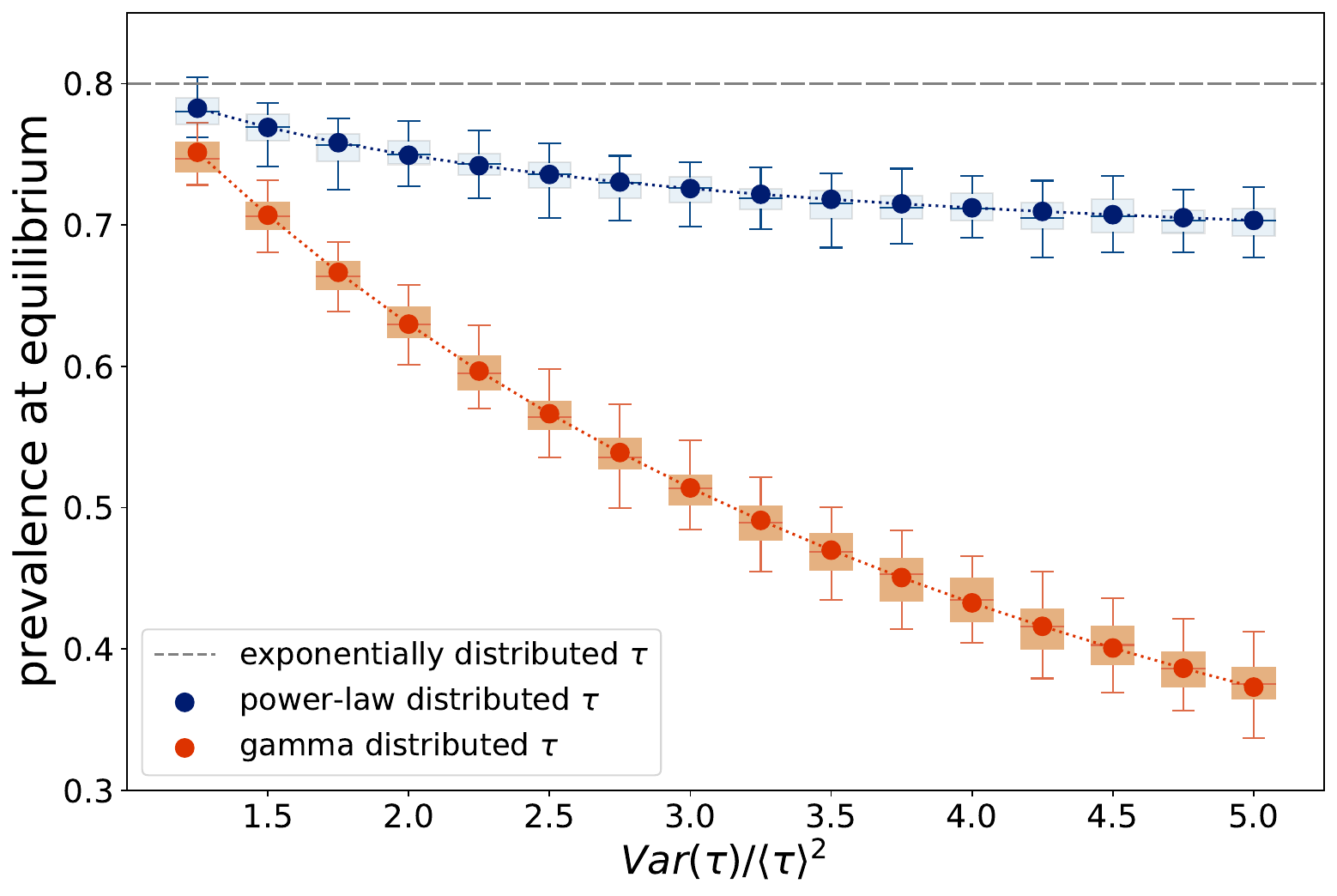}
\caption{Endemic prevalence in the {\itshape host-based} scheme as a function of the relative variance in $g(\tau)$. The gray line represents the endemic equilibrium for exponentially-distributed $\tau$ (constant $\mu$). Red and blue dots represent the numerical solution of Eq.~(\ref{eq:endemicL}), respectively for gamma-distributed and power-law-distributed $\tau$ (see also Tab.~\ref{table:conversion}). Box-plots represent values of the endemic equilibrium obtained from 100 SIS stochastic simulations on an Erdős–Rényi network of $N=10^{3}$ nodes and average degree 5. Disease parameters were fixed at: $\pan{\tau}=300$ time steps, $\lambda=5\lambda_{c}$.} 
\label{figure:host_based}
\end{figure*}

In Sec.~\ref{sec:epithreshold} we showed that the average infectious period $\pan{\tau}$ alone completely determines the epidemic threshold. Equation~(\ref{eq:endemicL}) instead shows that higher moments of $\tau$ have an impact on the endemic equilibrium, in the case of the \textit{host-based} scheme. In Fig.~\ref{figure:host_based} we keep $\pan{\tau}$ fixed, and explore different levels of dispersion around it in case of gamma-distributed, and power-law-distributed infectious period. Comparison with exponentially-distributed $\tau$ (at same $\pan{\tau}$) shows that in the \textit{host-based} scheme i) higher variance gives consistently lower endemic prevalence; ii) at fixed variance, gamma-distributed $\tau$ leads to lower prevalence than power-law-distributed $\tau$.

\section{APPLICATION TO MRSA DIFFUSION IN HEALTHCARE SETTINGS}
\label{sec:casestudy}

Methicillin-resistant Staphylococcus aureus (MRSA) is responsible for severe bacterial infections. Its acquired resistance to antimicrobial treatment makes it one of the most dreaded infections occurring in healthcare settings~\cite{hassoun_incidence_2017}. Patients can get colonized through direct contact with asymptomatic carriers (including healthcare workers). Outbreaks of MRSA infection increase mortality, hospitalization times, and are difficult and costly to contain~\cite{hassoun_incidence_2017}. 

\begin{figure*}[htbp]
    \centering
    \includegraphics[width=\linewidth]{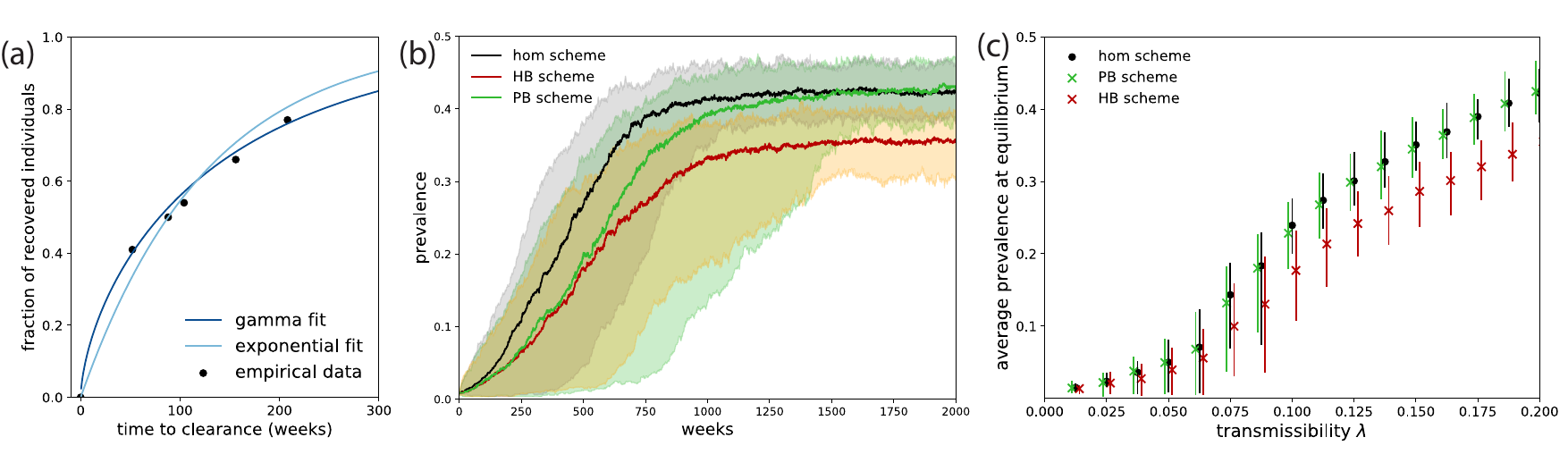}  
    \caption{MRSA infection spreading over a contact network within a hospital. (a) Gamma vs. exponential estimated cumulative distribution for MRSA colonization duration. Black dots represent empirical data on time to clearance from \cite{shenoy_natural_2014}. Values of parameters estimated at maximum likelihood: $r=0.007$, $\kappa=0.62$, $\theta=238.04$ (see Tab.~\ref{table:conversion} for the definition). AIC = 316 and 319 for gamma and exponential distribution respectively. (b) Colonization prevalence over time, for the {\itshape host-based} (red) and the {\itshape population-based} scheme (green) with a gamma-distributed infectious period, in comparison with the homogeneous scheme (black) assuming an exponential distribution and a constant recovery rate. Here $\lambda$ was fixed at $0.2$.
    (c) Colonization prevalence at equilibrium, for different values of transmissibility $\lambda$. Results are averaged over 100 stochastic runs. Uncertainty bars and shaded areas indicate 95\% probability ranges. 
    Parameters of the gamma-distributed $\tau$ are the fitted estimates. The rate of the exponential distribution is fixed to have the same $\pan{\tau}$ as the gamma-distributed one.
    Contact data are aggregated at a weekly time scale, getting a time-evolving, weighted network where weights encode contact duration over a specific week.
    }
    \label{figure:MRSA}
\end{figure*}

MRSA carriage duration is non-exponentially distributed. We analyzed data on time to observed clearance~\cite{shenoy_natural_2014} to reconstruct the distribution of carriage time $g(\tau)$. We fitted the data through maximum likelihood using exponential and Gamma distributions. The Akaike Information Criterion (AIC) selected the Gamma distribution as the best-fitting model (see Fig.~\ref{figure:MRSA}). This supports therefore the application of the approach presented here, as accurate model predictions are needed to improve surveillance and response against MRSA diffusion.

We use the estimated $g(\tau)$ to simulate the spread of MRSA carriage on a real network of contacts among $590$ individuals (both patients and healthcare workers) collected through wearable sensors in a long-term and rehabilitation facility in Northern France~\cite{obadia_detailed_2015, duval_measuring_2018}. We simulated both the {\itshape host-based}, and the {\itshape population-based} schemes. We also considered the scenario of constant recovery rate (exponentially-distributed $\tau$) as benchmark, with the same $\pan{\tau}$. The results of the simulations are displayed in Fig.~\ref{figure:MRSA}(b) and (c). 

Epidemic trajectories in Figure~\ref{figure:MRSA}(b) show that heterogeneity in the recovery rates has an effect in slowing down disease spread with respect to the homogeneous scheme. When approaching the endemic equilibrium, Figure~\ref{figure:MRSA}(c) confirms that the homogeneous modeling scheme with a fixed rate and the {\itshape population-based} scheme share the same endemic prevalence, even in the case of a realistic temporal contact network. Instead, in the {\itshape host-based} scheme the predicted endemic prevalence turns out to be smaller. As the transmission rate $\lambda$ decreases, the values for the three schemes converge, supporting the idea that they all hold the same epidemic threshold. 

\section{DISCUSSION}
\label{sec:conclusion}

A fixed recovery rate across individuals and throughout the epidemic outbreak fails in reproducing realistic distributions of infectious periods. Yet, it is the key modeling ingredient of traditional approaches because it treats recovery as a Markovian process, i.e. a spontaneous decay. This assumption allows analytical calculations, and largely simplifies numerical implementations. Based on a novel mapping of the infectious period distributions into a recovery rate distribution, we introduced a modeling framework that can capture and integrate arbitrary infectious period distributions that can be informed by empirical data, while remaining analytically and numerically treatable.

Collected empirical data usually provide information on the infectious period at the population level, not at the individual level, through a distribution of values. This 
lack of resolution at the host level opens the path to two possible modeling schemes,  assuming either an immutable recovery rate per host ({\itshape host-based} scheme), or a rate that can be updated at each infection episode because it is altered by factors affecting the immune response of the individual  ({\itshape population-based} scheme). When data on re-infection of the same individual are too scarce to estimate a distribution for each host (as it is often the case, with a few exceptions \cite{carrat_time_2008}), the two schemes become empirically equivalent, but they conceal significant differences in terms of predictions.

We analytically prove that the epidemic threshold, in the case of any generative network model for hosts interactions, does not depend on the scheme chosen. We also prove that such threshold depends only on the average infectious period,  making the standard assumption of constant recovery rate sufficient to correctly model the behavior around the disease-free state. Differences emerge when the system moves away from the critical point. The endemic disease prevalence --~a predictor of how easy it is to eradicate a disease in a population~-- is quantitatively different in each scheme, as shown in theoretical examples of disease spread and in a case study applied to the spread of the multiresistant bacteria MRSA in healthcare settings.

The difference extends to the out-of-equilibrium dynamics, as disease spreading in the {\itshape population-based} scheme is faster with respect to the {\itshape host-based} scheme.

Our findings show that modeling heterogeneity within the population-based or the host-based paradigms, although apparently equivalent, has a considerable impact on the endemic disease prevalence, and caution should be taken when addressing specific epidemic contexts where data do not allow to distinguish between the schemes. This problem might disappear naturally in contexts in which individual recovery rates have a concrete meaning or can be measured directly, e.g. in information diffusion processes \cite{kumar_information_2021}, but in the general case of biological diseases,
modelers should be aware that an arbitrary choice of the scheme may represent a potential source of bias to be considered.

\section*{ACKNOWLEDGMENTS}
This study was partially funded by ANR projects SPHINX (ANR-17-CE36-0008-05) and DATAREDUX (ANR-19-CE46-0008-03).

\renewcommand{\thesection}{\Alph{section}}
\titleformat{\section}[block]{\normalfont\fontsize{10}{15}\bfseries\filcenter}{APPENDIX \thesection :}{2.0mm}{}

\begin{appendices}

\section{Existence of the function $f$}
\label{sec:momentproblem}
The conditions on $g(\tau)$ under which there exists a probability density function $f(\mu)$ solving Eq.(\ref{eq:joint}) can be described in terms of a \textit{moment problem} \cite{schmudgen_moment_2017}. Let us evaluate the $n$-th derivative of $g(\tau)$ in $\tau=0$ from Eq.(\ref{eq:joint})
\begin{equation}
    \frac{\de{}^{n}}{\de{\tau^{n}}}g(\tau)|_{\tau=0}=(-1)^{n} \int_{0}^{+\infty} \de{\mu} f(\mu)\,\mu^{n+1}=(-1)^{n}m_{n+1}
\end{equation}
where $m_{n}$ indicates the $n$-th moment of $f(\mu)$. Thus a solution of Eq.(\ref{eq:joint}) exists if and only if the sequence
\begin{equation}
    \begin{aligned}
        m_{n+1}=& (-1)^{n}\frac{\de{}^{n}}{\de{\tau}^{n}}g(\tau)|_{\tau=0} \qquad \forall n\in \mathbb{N}\\
        m_0=&1
    \end{aligned}
\end{equation}
is a Stieljies moment sequence, i.e. it represents the sequence of moments of a measure on the interval $[0,+\infty)$. We set $m_0=1$ as we are looking for a probability measure. A sufficient and necessary condition for a real sequence $\{m_{n}\}_{n\in{\mathbb{N}}}$ to be a Stieljies moment sequence states that the Hankel matrices
\begin{equation}
    H^{(1)}_{n}=\begin{pmatrix}
    m_{0}&m_{1}&\dots&m_{n} \\
    m_{1}&m_{2}&\dots&m_{n+1}\\
    \vdots&\vdots&\ddots&\vdots\\
    m_{n}&m_{n+1}&\dots&m_{2n}
    \end{pmatrix}
\end{equation}
\begin{equation}
    H^{(2)}_{n}=\begin{pmatrix}
    m_{1}&m_{2}&\dots&m_{n+1} \\
    m_{2}&m_{3}&\dots&m_{n+2}\\
    \vdots&\vdots&\ddots&\vdots\\
    m_{n+1}&m_{n+2}&\dots&m_{2n+1}
    \end{pmatrix}
\end{equation}
need to be positive semi-definite for any $n\in\mathbb{N}$ \cite{schmudgen_moment_2017}. This property is useful to assess if the framework in terms of recovery rates is applicable or not given a certain $g(\tau)$. 

In Section~\ref{sec:mapping} we presented the analytical form of the distribution of recovery rates $f(\mu)$ when $g(\tau)$ is a $Gamma(\kappa,\theta)$ with $\kappa<1$. We can show that such distribution does not exist when $\kappa=2$. Indeed, for $\displaystyle{g(\tau)=\theta^{-2}\tau e^{-\tau/\theta}}$, the sequence turns out to be $m_{n}=-(n-1)\theta^{-n}$ and the Hankel matrix $H^{(1)}_{2}$ is not positive semi-definite.

\section{Epidemic threshold of the generic network model}
\label{sec:epithresholdApp}

The generic rank-$r$ network model is defined by its metadegrees ($r$ properties for each node), encoded in the $n\times r$ matrix $V$, and the signature of the nonsingular metric $\Delta$. See Ref.~\cite{valdano_exact_2019} for further details.
The adjacency matrix of such model is the rank-$r$ matrix $A=V\Delta V^T$. Let $M$ be a diagonal matrix containing the recovery rate $\mu_j$ of node $j$ in its $j$-th diagonal entry. Then, the linearized evolution of the disease close to the disease-free state follows the vector equation $\dot{\bf{x}} = (-M+\lambda V\Delta V^T)\textbf{x}$, where $x_j(t)$ is the probability that node $j$ is infectious at time $t$. Finding the epidemic threshold means finding the lowest value of $\lambda$ for which $-M+\lambda V\Delta V^T$ as a zero eigenvalue. We can compute the characteristic polynomial of this matrix using Ref.~\cite{golub_modified_1973}: $p(t) = \det \{\Delta^{-1}-\lambda V^{T}(t+M)^{-1}V\}$. The condition of the zero eigenvalue is then $\det \left\{1-\lambda V^{T}M^{-1}V \Delta \right\}=0$. The threshold condition that follows is $\lambda_c=1/\rho(V^{T}M^{-1}V\Delta)$, where $\rho$ is the spectral radius.

The last step consists in proving the following: $V^{T}M^{-1}V\Delta=\pan{\mu^{-1}}V^{T}V\Delta$. Let us call $Z=V^{T}M^{-1}V\Delta$ and compute its entry $Z_{\alpha\beta}$
\begin{equation*}
\begin{aligned}
    Z_{\alpha\beta}&=\sum_{i,j=1}^{N}\sum_{\gamma=1}^{r}V_{i\alpha}\delta_{ij}\mu_{i}^{-1}V_{j\gamma}\Delta_{\gamma\beta}=\\
    &=\sum_{\gamma=1}^{r}\Delta_{\gamma\beta}\sum_{i=1}^{N}V_{i\alpha}\mu_{i}^{-1}V_{i\gamma}=\\
    &=\sum_{\gamma=1}^{r}\Delta_{\gamma\beta}\sum_{i=1}^{N}(v_{\alpha})_{i}\mu_{i}^{-1}(v_{\gamma})_{i}=\\
    &=\sum_{\gamma=1}^{r}\Delta_{\gamma\beta} \pan{v_{\alpha}\mu^{-1}v_{\gamma}}N=\\
    &=\pan{\mu^{-1}}\sum_{\gamma=1}^{r}\Delta_{\gamma\beta}\pan{v_{\alpha}v_{\gamma}}N=\\
    &=\pan{\mu^{-1}}\sum_{\gamma=1}^{r}\sum_{i=1}^{N}V_{i\alpha}V_{i\gamma}\Delta_{\gamma\beta}\\
    &=\pan{\mu^{-1}}B_{\alpha\beta}
\end{aligned} 
\end{equation*}
under the assumption $\pan{v_{\alpha}\mu^{-1}v_{\gamma}}=\pan{\mu^{-1}}\pan{v_{\alpha}v_{\gamma}}$, where $v_{\alpha}$ is the $\alpha$-th column of the matrix $V$, i.e. the $\alpha$-th metadegree, and $B$ is the matrix $V^{T}V\Delta$, obtained from $A$ after rank reduction, following the notation in \cite{valdano_exact_2019}. As the matrix $A$ and $B$ share the spectral properties, we can conclude that
\begin{equation} 
    \lambda_{c}=\frac{1}{\rho(V^{T}M^{-1}V\Delta)}=\frac{1}{\pan{\mu^{-1}}\rho(B)}=\frac{1}{\pan{\tau}\rho(A)}.
\end{equation}

\section{Endemic equilibrium in the {\itshape population-based} scheme}
\label{sec:endeq}

In this section we derive the endemic prevalence in the {\itshape population-based} scheme and we show that it coincides with the one obtained when using a homogeneous recovery rate.
We consider a contact network with a fixed degree distribution $P(k)$, the so-called configuration network model \cite{newman_networks:_2010}. 

Within the homogeneous modeling scheme, a constant recovery rate $\mu$ is assigned to each individual in the population, so that the infectious period $\tau$ is exponentially distributed with mean $\pan{\tau}={\mu}^{-1}$.  Let $x_k(t)$ be the probability that a node with degree $k$ is infected at time $t$. According to the degree-based mean-field approximation \cite{boguna_epidemic_2002, barrat_dynamical_2008}, the equation describing the evolution of the SIS model is 
\begin{equation*}
    \frac{\de{x_k(t)}}{\de{t}} = -\mu x_k(t) + \lambda (1-x_k(t))k\sum_{k'} P(k'|k)x_{k'}(t)    
\end{equation*}
where $P(k'|k)$ is the probability that a node with degree $k$ is connected with a node of degree $k'$. In order to derive the number of infected individuals at equilibrium, one must solve for $I_k$ the following equation
\begin{equation}
    I_k = \frac{1}{\mu} \left[\frac{\lambda}{N} (N_k-I_k) k \sum_{k'} \frac{P(k'|k)}{P(k')}I_{k'}\right] 
\label{eq:endemic_class}
\end{equation}
where $I_{k}$ is the number of infected at equilibrium that have degree $k$. 

Now we assume the \textit{population-based scheme} and consider a general distribution $g(\tau)$ for the infectious period, and the corresponding distribution $f(\mu)$ for the recovery rates derived from Eq.~(\ref{eq:laplace}). Each individual is assigned a recovery rate $\mu_j$ from $f(\mu)$, updated by resampling after each recovery. We can still reason in terms of classes of degree $k$, but we need to further divide each class according to the recovery rate. Let us define $x_{j k}(t)$ as the probability that a node with recovery rate $\mu_j$ and degree $k$ is infected at time $t$. Then we can write
\begin{equation*}
\begin{aligned}
        \frac{\de{x_{j k}(t)}}{\de{t}} = &-\mu_j x_{j k}(t) +\\ &+\lambda (1-x_{j k}(t))k\sum_{k'}\sum_{h} P(\mu_h, k'|\mu_j, k)x_{h k'}(t)
\end{aligned}
\end{equation*}
where $P(\mu_h, k'|\mu_j, k)$ is the probability that a node with recovery rate $\mu_j$ and degree $k$ has a link to a node with recovery rate $\mu_h$ and degree $k'$. We assume that recovery rate and degree are uncorrelated, i.e. $P(\mu_h, k'|\mu_j, k) = P(\mu_h,k'|k) = P(\mu_h) P(k'|k)$
so we obtain
\begin{equation*}
    \frac{\de{x_{j k}(t)}}{\de{t}} = -\mu_j x_{j k}(t) + \lambda (1-x_{j k}(t))k\sum_{k'} P(k'|k) x_{k'}(t) 
\end{equation*}
since $\sum_{k'}P(k'|k)\sum_{h} P(\mu_h)x_{h k'}(t)=\sum_{k'} P(k'|k) x_{k'}(t)$. 
In terms of number of infected $I_{jk}$ with recovery rate $\mu_j$ and degree $k$, we obtain 
\begin{equation*}
    \frac{\de{I_{j k}(t)}}{\de{t}} = -\mu_j I_{j k}(t) + \frac{\lambda}{N} S_{jk}(t) k\sum_{k'} \frac{P(k'|k)}{P(k')} I_{k'}(t) 
\end{equation*}
The quantity $S_{jk}(t)$ is the number of susceptible nodes at time $t$ with recovery rate $\mu_j$ and degree $k$. This is equal to $S_k(t)$ multiplied by the probability of being assigned the recovery rate $\mu_j$ at the time of the last recovery, i.e.  $S_{jk}(t) = P(\mu_j)S_{k}(t)$. 
\begin{equation*}
    \frac{\de{I_{j k}(t)}}{\de{t}} = -\mu_j I_{j k}(t) + \frac{\lambda}{N} P(\mu_j)S_{k}(t) k\sum_{k'} \frac{P(k'|k)}{P(k')} I_{k'}(t) 
\end{equation*}
Solving for the equilibrium, and summing over the index $j$, we obtain
\begin{equation*}
\begin{aligned}
    I_k &= \sum_j I_{jk} = \\
    &=\sum_j \frac{P(\mu_j)}{\mu_j} \left[\frac{\lambda}{N}(N_k-I_{k})k\sum_{k'} \frac{P(k'|k)}{P(k')} I_{k'}\right]=\\
    &=\pan{\mu^{-1}} \left[\frac{\lambda}{N}(N_k-I_{k})k\sum_{k'} \frac{P(k'|k)}{P(k')} I_{k'}\right]\\
    \end{aligned}
\end{equation*}
which is equal to Eq.~(\ref{eq:endemic_class}) provided that the mean infectious period $\pan{\tau}=\pan{\mu^{-1}}$ is the same as the one assumed in the homogeneous modeling scheme. In conclusion, within the \textit{population-based} scheme, the endemic prevalence depends only on the average infectious period, and not on its distribution. 

\section{Endemic equilibrium in the {\itshape host-based} scheme}
\label{sec:endEqui}
In this section we derive the endemic prevalence in the {\itshape host-based} scheme, in the case of homogeneous mixing, as stated in the main text in Eq.~(\ref{eq:endemicL}).

Let $x_{j}(t)$ be the probability that an individual with recovery rate $\mu_j$ is infected at time $t$. Then,
\begin{equation}
    \frac{\de{}}{\de{t}} x_{j}(t) = -\mu_j x_{j}(t) + \frac{\lambda}{N} (1-x_{j}(t)) \sum_h x_h(t)
\end{equation}
where the average connectivity of the homogeneous network is absorbed in the transmission rate $\lambda$. By setting $\de x_{j}(t)/\de t=0$, we find 
\begin{equation}
    x_{j} = \lambda x_{eq}/(\mu_j +\lambda x_{eq})
\end{equation} 
where $x_{eq}$ is the endemic prevalence. In continuous terms, by integrating on both sides over all possible values of $\mu$, we get the following equation for the endemic prevalence $x_{eq}$:
\begin{equation}
        x_{eq} = \int_{0}^{\infty}\de{\mu} f(\mu) \left[ 1+\frac{\mu}{\lambda x_{eq}} \right]^{-1}
\label{eq:hb_equilibrium}
\end{equation}
We rewrite it as follows:
\begin{equation}
            1 = \lambda \int_{0}^{\infty}\de{\mu} f(\mu) \mu^{-1} \left[ 1+\frac{\lambda x_{eq}}{\mu} \right]^{-1}
\end{equation}
Expanding the inside of the integral as a geometric series, we get:
\begin{equation}
    \lambda \sum_{n=0}^\infty \left( -\lambda x_{eq} \right)^{n} \int_{0}^{\infty}\de{\mu} f(\mu) \mu^{-(n+1)} = 1.
\end{equation}
The integral in {\itshape lhs} is $\pan{\mu^{-(n+1)}}$, so we can use Eq.~(\ref{eq:relazMomenti}), re-index the sum, and get to
\begin{equation}
    \sum_{n=0}^\infty \left( -\lambda x_{eq} \right)^{n} \frac{\pan{\tau^n}}{n!} = 1-x_{eq}.
\end{equation}
Now the {\itshape lhs} is by definition the moment-generating function of $g$, evaluated in $-\lambda  x_{eq}$. Given that such argument is never positive, this is also --~by definition of moment-generating function~-- the Laplace transform of $g$, evaluated in $\lambda  x_{eq}$.
We thus get to the final form of the equation of the endemic equilibrium:
\begin{equation}
    \mathcal{L}\left[g\right]\left( \lambda x_{eq} \right) = 1-x_{eq}.
\end{equation}

\end{appendices}

\begin{table*}[htbp]
\centering
\begin{tabular}{lclcc}
\hline
\hline
\rule{0pt}{3ex}
infectious period distribution $g(\tau)$ &\qquad& recovery rate distribution $f(\mu)$&\qquad & $\pan{\tau}$\\
\hline
\rule{0pt}{4ex}
\small{Exponential} \hspace{2.0cm}$r\,e^{-r\tau}$ & & \small{Dirac delta} \hspace{2.3cm} $\delta(\mu-r)$&& $r^{-1}$\\
\rule{0pt}{4ex}
\small{Gamma($\kappa,\theta$)} \hspace{1.3cm} $\displaystyle\frac{\tau^{\kappa-1} e^{-\tau/\theta}}{\Gamma(\kappa)\theta^{\kappa}}$ & & \small{Power-law}\hspace{0.8cm}  $\displaystyle\frac{\sin(\kappa\pi)}{\pi\,\mu}\,(\theta\mu-1)^{-\kappa}H(\theta\mu-1)$&&$\kappa\theta$\\
\rule{0pt}{4ex}
\small{Power-law$(h,\epsilon)$} \hspace{0.5cm} $\displaystyle(h-1)\epsilon^{h-1}(\tau+\epsilon)^{-h}$ & & \small{Gamma($h-1,\epsilon^{-1}$)}\hspace{0.8cm}  $\displaystyle{\frac{\epsilon^{h-1}\mu^{h-2}\,e^{-\epsilon\mu}}{\Gamma(h-1)}}$\vspace{3.0mm}&&$\displaystyle{\frac{\epsilon}{(h-2)}}$\\ 
\hline
\hline
\end{tabular} 
\caption{Mapping from the infectious period distribution $g(\tau)$ to the recovery rate distribution $f(\mu)$, and corresponding average infectious period $\pan{\tau}$, for some commonly used infectious period distributions. $\delta$ is the Dirac delta distribution. $H$ is the Heaviside step function. For gamma-distributed $g$, the displayed $f$ is valid for $\kappa<1$. For power-law distributed $g$, $h>2$ ensures the existence of both $f$, and $\pan{\tau}$.}
\label{table:conversion} 
\end{table*}

\end{multicols}

\centering\rule{8cm}{0.4pt}
\begin{multicols}{2}
\printbibliography[heading=none] 
\end{multicols}

\end{document}